\documentclass[rmp,twocolumn]{revtex4}
\usepackage{graphicx}
\ifnum\lefthyphenmin<2\lefthyphenmin=2\fi
\ifnum\righthyphenmin<2\righthyphenmin=2\fi
\newcommand{\cQ}{{\cal Q}}

\begin{document}
\title{\bf Mechanically probing the folding pathway of single RNA 
molecules}
\author{Ulrich Gerland$^*$}
\author{Ralf Bundschuh$^\dagger$}
\author{Terence Hwa$^*$}
\affiliation{$^*$Department of Physics, University of California at San Diego, 
La Jolla, California 92093-0319}
\affiliation{$^\dagger$Department of Physics, The Ohio State University,
174 W 18th Av., Columbus, Ohio 43210-1106}
\date{\today}
%
% Abstract
%
\begin{abstract}
We study theoretically the denaturation of single RNA molecules by mechanical 
stretching, focusing on signatures of the (un)folding pathway in molecular 
fluctuations. 
Our model describes the interactions between nucleotides by incorporating 
the experimentally determined free energy rules for RNA secondary structure, 
while exterior single stranded regions are modeled as freely jointed chains. 
For exemplary RNA sequences (hairpins and the {\it Tetrahymena thermophila} 
group I intron), we compute the quasi-equilibrium fluctuations in the 
end-to-end distance as the molecule is unfolded by pulling on opposite ends. 
Unlike the average quasi-equilibrium force-extension curves, these 
fluctuations reveal clear signatures from the unfolding of individual 
structural elements.  
We find that the resolution of these signatures depends on the spring 
constant of the force-measuring device, with an optimal value intermediate 
between very rigid and very soft. 
We compare and relate our results to recent experiments by 
Liphardt {\it et al.} [{\it Science} 292, 733-737 (2001)].
\end{abstract}
%
%\keywords{RNA secondary structure formation, 
%          single-molecule experiments, force-extension curve, 
%          statistical mechanics, polymer modeling}
%
\maketitle
The advent and refinement of techniques to apply and measure forces on 
single molecules has allowed remarkable experiments probing the elastic 
properties and structural transitions of biomolecules such as DNA 
\citep[see, e.g.,][]{smith,bockelmann,maier} and proteins 
\cite[see, e.g.,][]{rief}. 
Recently, single-molecule experiments using optical tweezers were also 
performed to study the unfolding of small RNA molecules under an applied 
force \cite{liphardt,liphardt02}. 
RNA molecules, besides functioning as messengers of sequence information 
in the process of protein synthesis, also have many biological functions 
that intricately depend on a precisely folded RNA structure, e.g., 
as ribosomal RNA or as self-splicing introns \cite{cech}. 
The formation of RNA structure (`RNA folding') is therefore an important 
biophysical process, which however is currently not understood in 
sufficient detail \cite{tinoco}. 
Pulling experiments \cite{liphardt,liphardt02}, possibly in combination with 
single-molecule fluorescence methods \cite{zhuang,zhuang02}, promise to 
reveal important aspects of RNA structure, its folding pathways and kinetics, 
and eventually its biological function. 

The experiment of \citet{liphardt} has shown that
pulling on simple structural units of RNA, e.g., a single hairpin,
yields characteristic features in the force-extension curve (FEC),
which can be used to deduce the unfolding free energy and the size of
the structural element.  Moreover, the observation of end-to-end
distance time-traces at constant force revealed that the unfolding and
refolding of these simple structures proceeds directly without
intermediates, and the opening/closing rates could be extracted from
the time-traces. From the theoretical side, it is interesting to ask
how far, in principle, the pulling approach could be pushed to study
larger RNA molecules and how the resolution of such approaches depends
on parameters of the experimental setup. 
Here, we expand our previous model for RNA pulling experiments \cite{us}, 
and address these questions.  
The model incorporates the experimentally known free energy rules for 
RNA secondary structure \cite{walter} and a polymer model for the elastic 
properties of single-stranded RNA (ssRNA), but neglects pseudoknots and 
tertiary interactions, the energetics of which are currently poorly
characterized.

Our model yields predictions for force-extension measurements, including
fluctuations, and the mechanical (un)folding pathway for any given RNA
sequence. 
Below, we begin with the P5ab hairpin used by \citet{liphardt} and 
demonstrate that our model yields a FEC which is in semi-quantitative 
agreement with the experimental curve.
This agreement gives us confidence that our model is sufficiently realistic 
to permit its use to explore general questions regarding sequence-dependent 
signatures in mechanical single-molecule experiments on RNA. 

We first address the question of intermediates in the unfolding
pathway and show that, according to our model, a small modification in
the sequence of the P5ab hairpin can change its two-state folding
behavior and introduce a locally dominant intermediate.  However,
whether this intermediate state can be observed through quasi-equilibrium 
fluctuation measurements critically depends on the experimental conditions: 
If the force-measuring device is a ``soft spring'', the hairpin unfolds 
without any visible intermediates. 
Only when the force is measured with a stiff spring, an intermediate state
can be observed.

We then consider a larger RNA molecule, the group~I intron of
{\it Tetrahymena thermophila} with a sequence of about 400 bases 
and a known secondary structure containing many individual
elements \cite{cech}. 
Previous theoretical work predicted that equilibrium FEC's of large RNA's
are smooth and display no secondary structure dependent features, due
to a compensation effect between individual structural elements
\cite{us}.  Here, we find that even in the absence of structure-based
features in the FEC, measurements of the equilibrium fluctuations 
of the entire molecule can still be useful to obtain
information on the (un)folding pathway.
Also, for this longer molecule a good choice of the
stiffness of the force-measuring device is important for the
determination of the pathway. 

Finally, we compare the mechanical unfolding process studied here with the 
more conventional thermal unfolding. 
Again using the group I intron as an example, we find that the individual 
structural elements display significantly sharper opening transitions for 
force-induced than for thermal denaturation within our model. 
Indeed, in UV absorption experiments \cite{banerjee}, thermal melting of 
the intron shows only one broad signature associated with the opening of 
the secondary structure. 
\section*{Model} 
\begin{figure}
  \includegraphics[width=8cm]{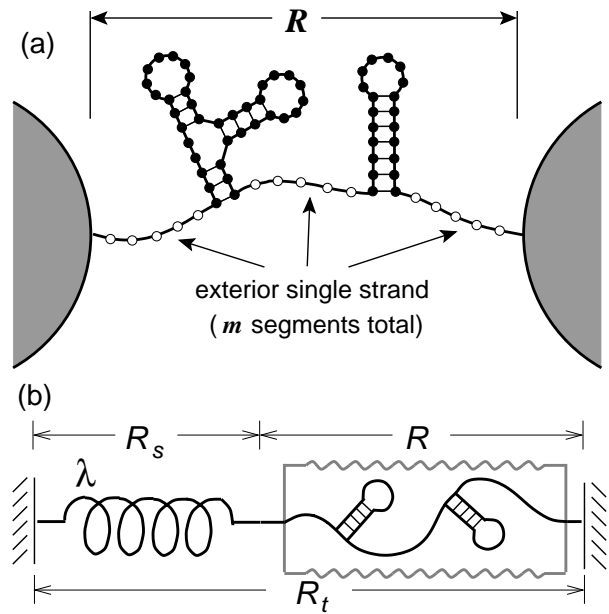}
  \caption{Sketch of the system considered here. 
  (a) In a typical experimental setup, the two ends of an RNA molecule 
  are attached to beads the position of which is controlled, e.g., with
  optical tweezers. 
  (b) In a theoretical model, the potential of the optical trap can be 
  approximated by a linear spring (with spring constant $\lambda$). 
  This spring is connected in series with a highly nonlinear elastic 
  element (the RNA molecule, usually with an additional linker). 
  The total extension, $R_t$, is externally controlled, while the 
  individual extensions, $R_s$ and $R$, undergo thermal fluctuations.}
  \label{figsprings} 
\end{figure}

We consider an experimental setup where the two ends of an RNA molecule 
are attached to a force- and extension-measuring device, e.g. an atomic 
force microscope or beads trapped by optical tweezers. 
For the sake of concreteness, we use the optical tweezer setup sketched in 
Fig.~\ref{figsprings}(a) as an example here. 
In an idealized theoretical model, the RNA molecule (in essence, a highly 
nonlinear elastic element for the present context) is connected in series 
with a linear harmonic spring, and the total extension is externally
controlled, see Fig.~\ref{figsprings}(b).
The linear spring models the force-measuring device. E.g., for optical 
tweezers, its harmonic potential approximates the potential of the optical 
trap, and the spring constant $\lambda$ is adjustable in the experiment 
through the laser intensity. 
The extension of the spring, $R_s$, corresponds to the position of the bead 
with respect to the minimum of the trapping potential, and is measured as 
a function of the total extension $R_t$. 
The average force acting on the RNA molecule and its average extension are 
\begin{eqnarray}
\label{force}
  \langle f \rangle &=& \lambda \, \langle R_s \rangle \;, \\
\label{extension}
  \langle R \rangle &=& R_t - \langle R_s \rangle \;.
\end{eqnarray}
Here, $\langle\ldots\rangle$ denotes a thermal average over all accessible 
conformations of the RNA molecule and the spring at fixed total 
extension $R_t$. Throughout this article, we assume that the pulling 
experiment is performed slowly enough, such that the RNA molecule can 
fully sample its conformational space on timescales which are short compared 
to the external timescale of the imposed stretching process. 
This limit corresponds to the quasi-equilibrium regime, where the work 
dissipated in the pulling process is negligible. 
The average $\langle\ldots\rangle$ then includes an average over all 
possible secondary structures of the molecule.

{\bf Partition function.} 
To calculate the averages in Eqs.~(\ref{force}) and~(\ref{extension}), as well
as the averages of other observables to be introduced below, we first determine
the partition function, $Z(R_t)$, of the RNA molecule and the spring at 
fixed total extension $R_t$. 
In previous work \cite{us}, we computed the partition function of the RNA 
molecule alone in the {\em fixed distance ensemble}. 
Here, we explicitly take the force-measuring spring into account; 
varying the spring constant $\lambda$ interpolates between the fixed 
distance ensemble for stiff springs (large $\lambda$) and the fixed force 
ensemble for soft springs (small $\lambda$). 
Intermediate spring constants effectively put the RNA molecule into a 
{\em mixed ensemble}. 
The explicit consideration of the spring in the model is necessary here, not 
only for a closer modeling of the actual experimental setup, but also from 
a theoretical point of view, since the force {\em fluctuations} diverge in 
the fixed distance ensemble and our aim is to study fluctuation 
measurements. 

A model for an experimental setup such as the one sketched in 
Fig.~\ref{figsprings}(a) requires two separate parts, one that contains 
only information on the free energies for the secondary structures of the 
RNA sequence, and another which describes the polymer properties of 
ssRNA \cite{us}. 
The coupling between these two parts is through the total length of the 
{\em exterior} ssRNA segments of the molecule, i.e. the number $m$ of bases 
that actually ``feel'' the applied force, see Fig.~\ref{figsprings}(a). 
For every fixed total extension $R_t$, the system has to find a compromise 
between lowering the RNA binding energy by {\em decreasing} $m$, and 
gaining entropy (plus lowering elastic energy) by {\em increasing} $m$. 
The secondary structure part is given by the partition function, $\cQ(m)$, 
summed over all secondary structures of the RNA molecule with 
a fixed number $m$ of exterior single-stranded bases\footnote{We account for 
the width of stems within the exterior single strand by increasing the 
base count $m$ by three for each stem.}. 
The polymer properties of ssRNA then enter through a function 
$W_{\rm tot}(R_t;m)$, which denotes the {\em total} end-to-end 
distance distribution of a ssRNA molecule of $m$ bases in 
series with the spring\footnote{For simplicity, we restrict the spring to be 
collinear with the end-to-end distance of the RNA molecule.}. 
The total partition function, $Z(R_t)$ is then simply the convolution of 
$\cQ(m)$ with $W_{\rm tot}(R_t;m)$, 
\begin{equation}
  \label{part_funct}
  Z(R_t)=\sum_m \,\cQ(m)\,W_{\rm tot}(R_t;m)\;.
\end{equation}

The detailed calculation of $\cQ(m)$ is described in \citet{us}. 
Briefly, the calculation takes the experimentally determined rules 
for the binding free energies of RNA secondary structures \cite{walter} 
into account, but neglects tertiary structure effects and pseudoknots. 
Our method is based on a recursive method to calculate the partition 
function \cite{mccaskill} as implemented in the `Vienna RNA package' 
\cite{vienna}, but introduces an additional recursion relation for the 
calculation of $\cQ(m)$. 

The total end-to-end distance distribution of a ssRNA molecule of $m$ 
bases in series with the spring can be expressed as 
\begin{equation}
  \label{W_tot}
  W_{\rm tot}(R_t;m)=\int\limits_0^{\infty}{\rm d}R\;W_{\rm RNA}(R;m)\,
  \frac{e^{-\beta\lambda(R_t-R)^2/2}}{\sqrt{2\pi/\beta\lambda}}\;,
\end{equation}
where $W_{\rm RNA}(R;m)$ denotes the distribution of the ssRNA
molecule alone. We model the ssRNA as an elastic freely jointed chain,
i.e., as a chain of segments with unrestricted relative orientations, 
whereas the segment length $r$ is constrained by a harmonic potential 
$V(r)=\kappa(r-b)^2/2$. The average segment length $b$
corresponds to the Kuhn length of a non-interacting ssRNA chain. The
number of segments of the elastic freely jointed chain used to
represent an RNA molecule with $m$ bases is chosen as $ml/b$ where $l$
is the base to base distance of ssRNA. This yields $W_{\rm
RNA}(R;m)\approx C\frac{h}{2\pi R}\left[q(h)\right]^{ml/b}e^{-hR}$,
where $C$ is a normalization constant, $q(h)=\langle e^{-h{\bf \hat
z}\cdot{\bf r}} \rangle$, and $h$ is determined from
$R=\frac{mb}{l}\;\frac{\partial}{\partial h} \log q(h)$~\cite{us}.
Such a model has been shown to describe the elastic properties of
ssDNA molecules \cite{maier,mezard}.  Since we are not aware of the
corresponding data for RNA, we use the DNA values for the polymer
parameters as obtained by \citet{mezard} through fitting to the experiment
of \citet{maier}, i.e.  $l\!=\!0.7$nm, $b\!=\!1.9$nm, and
$(\kappa/k_B T)^{-1/2}\!=\!0.1$nm (we do not expect a large difference
in the single strand properties of DNA and RNA, because of the high
similarity between their chemical structures).  We take the free
energy parameters for RNA secondary structure as supplied with the
Vienna package (version 1.3.1) at room temperature $T=25^oC$.  The
salt concentrations at which the free energy parameters were measured
are $[\rm{Na}^+]=1M$ and $[\rm{Mg}^{++}]=0M$.

{\bf Observables.} 
{From} the partition function (\ref{part_funct}) we get the total free 
energy $F(R_t)=-k_BT \log Z(R_t)$, the average force 
$\langle f \rangle=\partial F(R_t)/\partial R_t$, and through 
Eqs.~(\ref{force}) and~(\ref{extension}) also the average extension 
$\langle R \rangle$ of the RNA molecule. 
The equilibrium fluctuations in the extension of the RNA molecule 
are given by 
$(\delta R)^2=\langle R^2\rangle - \langle R\rangle^2=
\big[\lambda-\partial^2\!F/\partial R_t^2\big]/\beta\lambda^2$. 
Using these relations, one can 
show\footnote{The derivation is straightforward after observing that 
$$\frac{\partial\langle f\rangle}{\partial \langle R\rangle}=
\frac{\partial\langle f\rangle}{\partial R_t}\Big/
\frac{\partial\langle R\rangle}{\partial R_t}=
\frac{\partial^2F(R_t)/\partial R_t^2}{1-(1/\lambda)
\partial^2F(R_t)/\partial R_t^2}\;.$$} 
that the variance is related to 
the derivative of the force-extension curve as measured 
{\em in the mixed ensemble} through 
\begin{equation}
  \label{varR}
    (\delta R)^2=\frac{k_BT}{\lambda+
    \partial\langle f\rangle/\partial\langle R\rangle}\;. 
\end{equation}
This exact relation simply expresses the intuitive result that the 
extension of the molecule undergoes large thermal fluctuations in the 
flat regions of the force-extension curve. 

To study the unfolding pathway, we introduce the weight $x_m(R_t)$ 
for the secondary structures with $m$ exterior open bases at a given 
value of $R_t$, 
\begin{equation}
  \label{xm}
  x_m(R_t)=\frac{\cQ(m)\,W_{\rm tot}(R_t;m)}{Z(R_t)}\;,
\end{equation}
which satisfy the normalization condition $\sum_m x_m=1$. 
The binding probability of base $i$ and $j$ at given $R_t$ may then be 
expressed as 
\begin{equation}
  \label{Pij}
  P_{ij}(R_t)=\sum_m x_m(R_t)\,p_{ij}(m)\;,
\end{equation}
where $p_{ij}(m)$ denotes the base pairing probability in the ensemble 
of structures with fixed $m$, which can be calculated as described 
previously \cite{us}. 
\section*{Results} 
\begin{figure}
  \includegraphics[width=8.5cm]{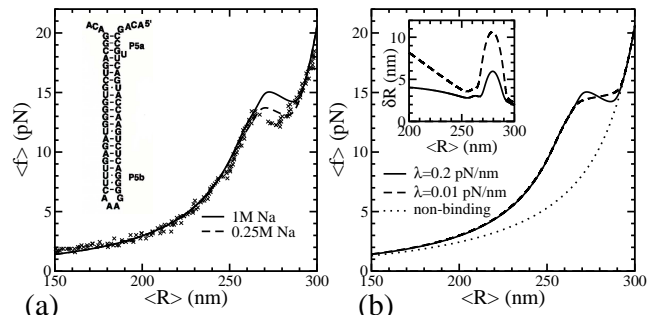}
  \caption{\label{P5ab_FEC}
  Force-extension curve (FEC) for the P5ab hairpin with added
  linker. (a) Native structure of the hairpin, its FEC as 
  experimentally determined by \citet{liphardt}, and the FEC
  resulting from our model for two different salt concentrations. 
  (b) FEC for P5ab as results from our model for two different spring 
  constants \protect$\lambda$ 
  (solid and dashed line; see main text for details). 
  The dotted line represents the FEC for a non-binding sequence of the 
  same length as P5ab. Inset: End-to-end distance fluctuations as a 
  function of the average extension.} 
\end{figure}

We first consider the P5ab hairpin, which is the simplest RNA molecule
studied in the experiment of \citet{liphardt}, and is of a size that
typically occurs as a structural subunit in larger RNA's. 
Fig.~\ref{P5ab_FEC}(a) shows its native structure and the experimental 
FEC for P5ab with added linker, taken with permission from
\citet{liphardt}. 
The superimposed solid line shows the FEC that results from our model 
as described above\footnote{The native structure of P5ab contains two 
non-standard G--A base pairs for which no quantitative information on 
the binding energies is available. We therefore follow the suggestion of
Ref.~\cite{liphardt} and approximate the binding energy of G--A
basepairs by replacing them with G--U pairs.} with a spring constant 
of $\lambda=0.2$~pN/nm. 
Here, we modeled the RNA-DNA hybrid linker, which in the experiment 
connects the RNA molecule to the beads, as a worm-like chain (WLC) with 
an experimentally known length of $L=320$nm. 
We used the worm-like chain interpolation formula \cite{bustamante}
\begin{equation}
  \label{WLC}
  f_{\rm WLC}(R)=\frac{k_BT}{l_p}\left(\frac{1}{4(1-R/L)^2}+\frac{R}{L}-
                 \frac{1}{4}\right)\;,
\end{equation}
where the value for the persistence length, $l_{p}=3.57$nm and the 
(experimentally unknown) origin of the distance scale were obtained 
by fitting to the experimental FEC at low forces. 
The shape of the experimental FEC is well captured by the theoretical 
curve, with a characteristic ``hump'' indicating the opening of the hairpin 
(the force drop in the hump occurs, because the released single-strand 
creates ``slack'' in the polymer). 
However, the force at which the hairpin opens is overestimated by the 
theoretical model. 
To assess the origin of this discrepancy, we applied a simple salt 
correction to the RNA binding (free) energies \cite{santalucia} to account 
for the different ionic conditions between the single-molecule experiment 
($[{\rm Na^+}]=0.25$M) and the standard conditions ($[{\rm Na^+}]=1.0$M). 
The resulting FEC (dashed line in Fig.~\ref{P5ab_FEC}a) yields 
near-quantitative agreement with the experiment, indicating that the 
difference in ionic conditions accounts for most of the discrepancy. 
The remaining discrepancy could be caused, for instance, by the freely 
jointed chain model for ssRNA or the unknown stacking energies for G--A 
basepairs. However, the good agreement we have achieved seems sufficient 
to justify the use of our model as a tool to study the general questions 
outlined in the introduction (in the following, we keep the salt 
concentration at the standard conditions).

The FEC's in Fig.~\ref{P5ab_FEC} contain useful information on the RNA 
molecule. For instance, the total binding free energy and the size of the 
hairpin can be read off by comparing to the FEC for a non-binding control 
sequence. 
Within our theoretical model, such control curves are obtained through a 
convolution of the WLC model (\ref{WLC}) for the linker with the 
freely-jointed chain model for the single-stranded RNA. 
The resulting curves could also be used as a reference for the experimental 
data, although ideally the control curve should be measured, using e.g. an 
oligonucleotide with a homogeneous sequence (note that it suffices to record 
the control for a single sequence length, since it can be straightforwardly 
rescaled to an arbitrary length). 
Fig.~\ref{P5ab_FEC}(b) shows our theoretical FEC for P5ab both for a stiff 
spring ($\lambda=0.2$~pN/nm, solid line) and a soft spring 
($\lambda=0.01$~pN/nm, dashed line), together with the control FEC (dotted) 
for a sequence of commensurate length. 
Since stretching the control sequence requires only work against entropic 
and elastic forces, while stretching of the P5ab hairpin requires additional 
work to break the basepairs, the area between the two FEC's equals the total 
binding free energy of P5ab. 
As Fig.~\ref{P5ab_FEC}(b) shows, not only the region of the P5ab FEC
where the basepairs are opened (the ``hump'' region) contributes to
this area, but there is also a significant contribution from the
initial part of the FEC.  The initial part of the P5ab FEC corresponds
to stretching the exterior single-stranded chain of the molecule,
which is shorter than the full molecule and hence has higher
entropic/elastic forces at the same extension. 
However, this additional entropic/elastic energy is
``stored'' in the molecule and is fully released when the hairpin is
(adiabatically) unzipped, i.e. in the ``hump'' region.  Note that as
long as the experiment is performed in quasi-equilibrium, the area
under the FEC is independent of the spring constant, and therefore the
same in the fixed-force and the fixed-distance ensemble.  Loosely
speaking, the FEC in the fixed-force ensemble could therefore be
considered as the `Maxwell construction' of the FEC in the
fixed-distance ensemble.

We now seek to characterize the unfolding process of the P5ab hairpin within 
our model. In the experiment of \citet{liphardt}, the time traces of the
end-to-end extension at fixed forces around 14 pN show the characteristic 
pattern of a bistable system, i.e. the unfolding proceeds directly, without 
intermediates. 
To calculate theoretical time traces would require a full kinetic treatment 
of the RNA folding/unfolding process which is beyond the scope of the 
present article. 
Instead, we determine the {\em amplitude} of the equilibrium 
fluctuations in the end-to-end extension, $\delta R$, as a function of the 
mean force or extension. 
Experimentally, this amplitude would be determined as a time-average, 
${\delta R}^2=T^{-1}\int_0^T\!dt\,(R(t)-\overline{R}\,)^2$, where $R(t)$ 
denotes the recorded time trace of the end-to-end extension and 
$\overline{R}$ its average. 
Our theoretical prediction is for the infinite time limit, $T\to\infty$, 
of this average. 
Note that the exact relation Eq.~(\ref{varR}) directly links the 
equilibrium fluctuations to the derivative of the averaged FEC. 
In practice, where the averaging period $T$ is finite and limited to a few 
minutes by instrumental drift \cite{liphardt}, the amplitude $\delta R$ can 
be determined with much greater accuracy than the derivative of the 
time-averaged FEC, since taking the derivative strongly amplifies the 
statistical error. 
Therefore, the quasi-equilibrium FEC and the equilibrium fluctuations, 
$\delta R$, can effectively be regarded as two independent and 
complementary sources of information. 

The inset of Fig.~\ref{P5ab_FEC}(b) shows the fluctuations $\delta R$ 
as a function of the average extension, 
$\langle R\rangle$, in the range of extensions over which the hairpin opens. 
We observe a single peak of the fluctuations at the force/extension 
where the transition takes place, which is consistent with the two-state 
behavior found experimentally. 
Physically, this peak is caused by continual kinetic fluctuations 
between the open and the closed state of the hairpin. 

\begin{figure}
  \includegraphics[width=8.5cm]{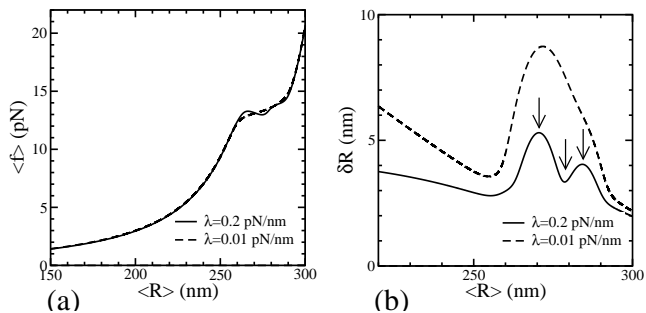} 
  \caption{\label{P5ab_uu}
  Calculated force-extension characteristics for the modified P5ab 
  hairpin (G--A basepairs replaced by A--A) for a hard 
  (\protect$\lambda=0.2$~pN/nm) and a soft (\protect$\lambda=0.01$~pN/nm) 
  harmonic potential of the optical tweezer. 
  (a) The average FEC, which resembles very much the FEC in 
  Fig.~\protect\ref{P5ab_FEC}(b).
  (b) The fluctuations of the extension as a function of the
  average extension. While the curve is very similar to the inset of
  Fig.~\protect\ref{P5ab_FEC}(b) for a soft spring, it develops a
  pronounced minimum corresponding to an intermediate state for a
  hard spring. The arrows indicate the positions at which the 
  $x_m$-distributions shown in Fig.~\ref{P5ab_mdistr} were calculated.}
\end{figure} 

\begin{figure}
  \includegraphics[width=8cm]{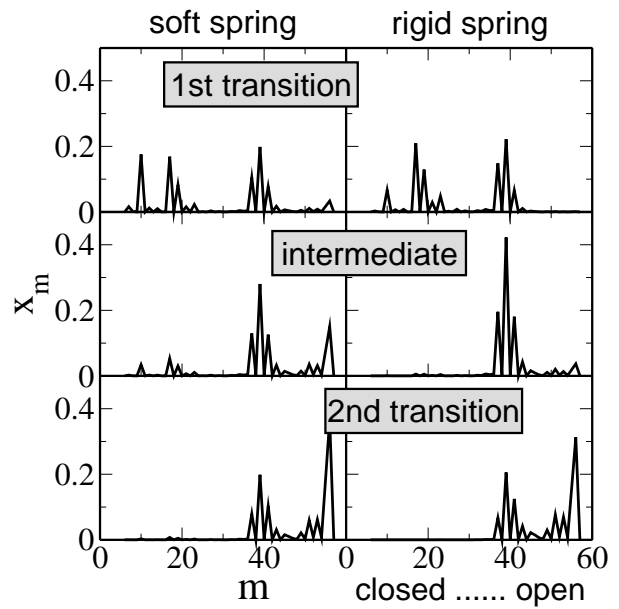}
  \caption{\label{P5ab_mdistr} 
  Probability distributions in $m$-space for the modified P5ab hairpin. 
  The left column is calculated with a soft spring ($\lambda=0.01$~pN/nm), 
  and the right column with a stiff spring ($\lambda=0.2$~pN/nm). 
  The curves in the top/middle/bottom row are calculated for different 
  average extensions $\langle R\rangle$ corresponding to the positions 
  indicated by the arrows in Fig.~\protect\ref{P5ab_uu}(b) 
  [first peak, minimum, and second peak in the fluctuation curve, 
  respectively]. 
  For the stiff spring, the distribution at the minimum is almost entirely 
  localized on the intermediate state. 
  In contrast, the distribution is spread out over all three states for the 
  soft spring at the same average extension.} 
\end{figure}

{\bf Hairpin with intermediate.}
Next, we address the question whether the simple two-state behavior 
observed for P5ab is a generic property for hairpins of this typical size. 
The discussion of this question also serves as a preparation for our 
ensuing study of a larger RNA with more complicated structure. 
To this end, we modify the sequence of the P5ab hairpin slightly by 
replacing the two G--A basepairs with U's on both strands, which leads 
to a small interior loop in the hairpin. 
This change does not significantly affect the FEC, see 
Fig.~\ref{P5ab_uu}(a), however the fluctuations $\delta R$ take on a 
qualitatively different behavior, as shown in Fig.~\ref{P5ab_uu}(b): 
For a soft spring, the $\delta R$-curve shows only a single peak as before, 
but a stiff spring yields two maxima with a pronounced minimum in between.
At this minimum, the configurational distribution of the molecule is 
localized on an intermediate state, which strongly reduces the fluctuations 
in the extension. 
This is demonstrated explicitly in Fig.~\ref{P5ab_mdistr}, which shows the 
probability distribution in $m$-space, i.e. $x_m$ as given by Eq.~(\ref{xm}) 
(see caption for details). 

The stiffness of the external spring therefore limits the attainable 
resolution of the measurements: At low resolution the hairpin appears as 
a two-state folder, while an intermediate state appears at higher 
resolution. This observation is in line with the comment of 
\citet{fernandez} who discuss the experiment of 
\citet{liphardt} by analogy to patch clamp experiments, where an 
intermediate state in the opening of an ion channel appears only at 
high resolution. 
While an increase in the spring constant increases the resolution, 
it decreases the amplitude of the fluctuations, which renders them harder 
to detect. 
For the present purpose, the optimal value for $\lambda$ results from a 
tradeoff between low resolution and low amplitude. 
Clearly, increasing $\lambda$ can increase the resolution only up to the 
point where the floppiness of the RNA plus linker becomes resolution 
limiting (this point corresponds approximately to $\lambda=0.5$~pN/nm for 
the hairpin sequences with linker discussed above). 
This point marks the optimal choice for $\lambda$, unless the spacial 
resolution of the experimental apparatus is restrictive already at smaller 
values of $\lambda$. 
The role of the spring constant in dynamic force spectroscopy measurements 
is discussed by \citet{grubmueller}. 

\begin{figure}
  \includegraphics[width=8.0cm]{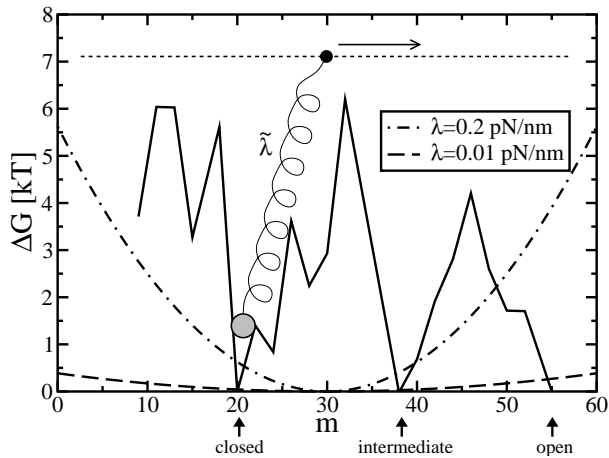}
  \caption{\label{stickslip} 
  Particle with an attached spring in the free energy landscape of the 
  modified P5ab hairpin. The other end position of the spring is 
  externally controlled \citep[see][for the details of the mapping to the RNA 
  unzipping problem]{us}. 
  The spring in this picture represents the measurement device, the linker, 
  and the ssRNA and has an effective spring constant $\tilde{\lambda}$. 
  Its harmonic potential for \protect$\lambda=0.01$~pN/nm
  and \protect$\lambda=0.2$~pN/nm is shown as the dashed and dash-dotted 
  lines, respectively. 
  With the soft spring,
  the particle can jump from the closed to the open state immediately as
  soon as the force is strong enough to overcome the energy barrier. The
  hard spring has a steeper harmonic potential and thus the particle
  will first jump into the intermediate state and later into the open
  state as the end of the spring is moved towards larger \protect$m$.
  }
%
% The two parabolas in this figure actually include two "experimental"
% parameters. First, it is to be noted that the spring in the figure really
% represents the combined spring of the trap (described by the real lambda)
% and the polymer spring of the single-stranded RNA. This combined
% lambda is of the order of 0.14 if the original lambda is 0.2. Thus,
% I used 0.14 for the lambda=0.2 curve. At lambda=0.01, the polymer
% spring does not play any role any more, since the external spring is
% so much softer that it dominates. Also, the length of a base was
% taken to be 0.6nm in order to convert the lambdas into m-space.
% While a fully stretched base has a length of 0.7nm in our model,
% the 0.6nm are an empirical number from our simulations correct at
% the relevant forces.
%
\end{figure} 

Intuitively, the fact that the appearance of the intermediate state 
depends on the spring constant is best understood by mapping the 
unzipping problem onto the problem of a particle with an attached 
spring in a random potential and coupled to a thermal bath 
(see Fig.~\ref{stickslip}). 
As the other end of the spring is moved in one direction, the particle 
performs a `stick-slip' motion \cite{bockelmann}. 
With a stiff spring, the particle hops over short distances and tends to 
be localized, while the particle can make long jumps, sliding over many 
valleys, when the spring is soft.

\begin{figure}
  \includegraphics[width=8.5cm]{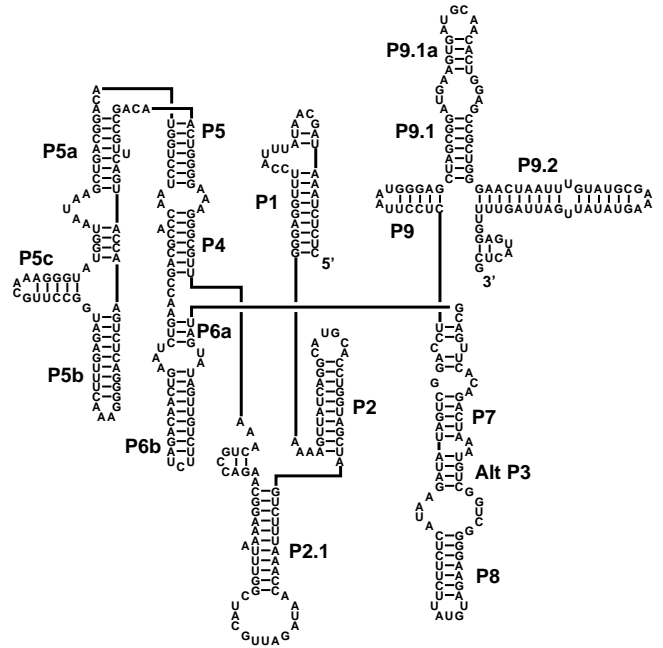}
  \caption{Secondary structure of the {\it Tetrahymena thermophila} 
  group I intron as obtained from the Vienna package. 
  The overall structure, and almost all individual basepairs, are identical to 
  the one of the inactive conformation as determined by \citet{pan} 
  (we follow the labeling introduced there).}
  \label{figstruct} 
\end{figure}

{\bf RNA molecule with multiple structural subunits.}
Longer RNA molecules typically have a complex structure with many 
structural subunits interacting through weak tertiary contacts \cite{tinoco}. 
A typical example for a structural subunit is the P5ab hairpin studied above, 
which is indeed extracted from a larger RNA molecule, the self-splicing 
intron of {\it Tetrahymena thermophila} with a sequence of 419 bases 
(Genbank \# J01235). 
The known secondary structure of this intron comprises 19 structural 
elements labeled P1, P2, P2.1, P3, etc. 
It was the first RNA molecule to be shown to have a catalytic activity 
and has since been studied in great detail \cite{cech}. 
Its active, i.e. self-splicing, conformation contains a pseudoknot, which 
is essential for the function. 
However, it also has a known stable and well characterized inactive 
conformation without pseudoknot \cite{pan,zhuang}. 
The minimum free energy structure obtained from the Vienna RNA Package 
(which supplies the basis to our model) is shown in Fig.~\ref{figstruct}. 
The overall structure, and almost all individual basepairs, are identical to 
the one of the inactive conformation as determined by \citet{pan} 
(we follow the labeling introduced there), which confirms 
that the RNA free energy rules describe the nucleotide interactions 
quite well. 
For the cyclized form of this intron, a very similar secondary structure 
was obtained by \citet{jaeger}. 

For complex secondary structures such as the one shown in
Fig.~\ref{figstruct}, it is interesting to study the physical process
of structure formation, whereby e.g. a newly produced RNA molecule
folds into its preferred structure(s).  For instance, one may ask, in
which order the structural elements in Fig.~\ref{figstruct} form,
i.e. what is the folding pathway? 
Mechanical single-molecule experiments offer a very controlled way to 
probe the folding process. 
In the quasi-equilibrium limit that we consider, the mechanical folding 
pathway is independent of (i) the direction, i.e. unfolding and refolding 
pathway are identical, and (ii) specific experimental parameters such as 
the spring constant $\lambda$. 
The latter is clear from the fact that the number of exterior unpaired 
bases, $m$, is a natural and well-defined reaction coordinate in the 
quasi-equilibrium limit. 
On average, the distribution $x_m$ will always shift to larger $m$ when 
$R_t$ is increased, and hence $\lambda$ can only affect the resolution of 
individual states in $m$-space, but not the average order in which they 
are visited. 

How can one determine the (un)folding pathway in a single-molecule 
experiment? As we have seen, pulling on an isolated subunit produces a 
characteristic signature in the FEC, see Fig.~\ref{P5ab_FEC}. 
If every structural subunit in a larger RNA molecule would produce such a 
characteristic signature, one could directly read off the folding pathway 
from the FEC. However, our previous theoretical study predicted that the 
quasi-equilibrium FEC of RNA's with many subunits will typically show no 
distinguishable signatures of individual subunits opening \cite{us}. 
This prediction is consistent with the experimental observation of smooth 
FEC's for long ssDNA molecules \cite{maier}. 
Three physical effects are responsible for the smoothness: 
(i) the `floppiness' of the external single-strand and the linkers, 
(ii) thermal fluctuations in the secondary structure, i.e. the contribution 
of suboptimal structures, and most importantly, 
(iii) the fact that changes in the extension of individual structural 
subunits can compensate each other, since only the total end-to-end distance 
is measured \cite{us}. 
As a result, quasi-equilibrium FEC's cannot be used to study the folding 
pathway (or the secondary structure) of larger RNA molecules. 

The smoothing mechanisms (ii) and (iii) can be strongly suppressed by 
performing the experiments at large pulling speeds which do not leave 
sufficient time for the molecule to sample the entire ensemble of different 
secondary structures with similar number of exterior unpaired bases, $m$, 
and free energy of folding. Hence, nonequilibrium FEC's of larger 
RNA molecules should be very rugged and display signatures of individual 
structural elements opening; see also the discussion in \citet{us}.
This was indeed observed recently in the Berkeley group 
(S. Dumont \& I. Tinoco Jr., private communication). 
{From} the theoretical side, the calculation of nonequilibrium FEC's is 
challenging, since the simple particle-in-a-landscape picture of 
Fig.~\ref{stickslip} breaks down in nonequilibrium: the free energy 
landscape is no longer well-defined. The secondary structures that are 
accessible for a molecule at a given extension and within a given time window 
depend on the present structure of the molecule and the detailed folding 
kinetics. While important steps towards a kinetic theory of RNA folding have 
been taken \cite{herve}, there is currently insufficient experimental 
information available to construct a full theoretical model. 

Here, we consider instead the equilibrium fluctuations in the end-to-end 
distance of the entire intron, and explore, within our theoretical model, 
how much information on the folding pathway can be obtained from this. 
While looking at the equilibrium fluctuations does not 
suppress any of the above-mentioned smoothing mechanisms, it does provide a 
greater resolution for the observation of structural transitions, as 
already discussed. We will see in the following that this enhanced 
resolution, which is dependent on the spring constant, is sufficient to 
reveal major steps of the unfolding process. 

\begin{figure}
  \includegraphics[width=7.5cm]{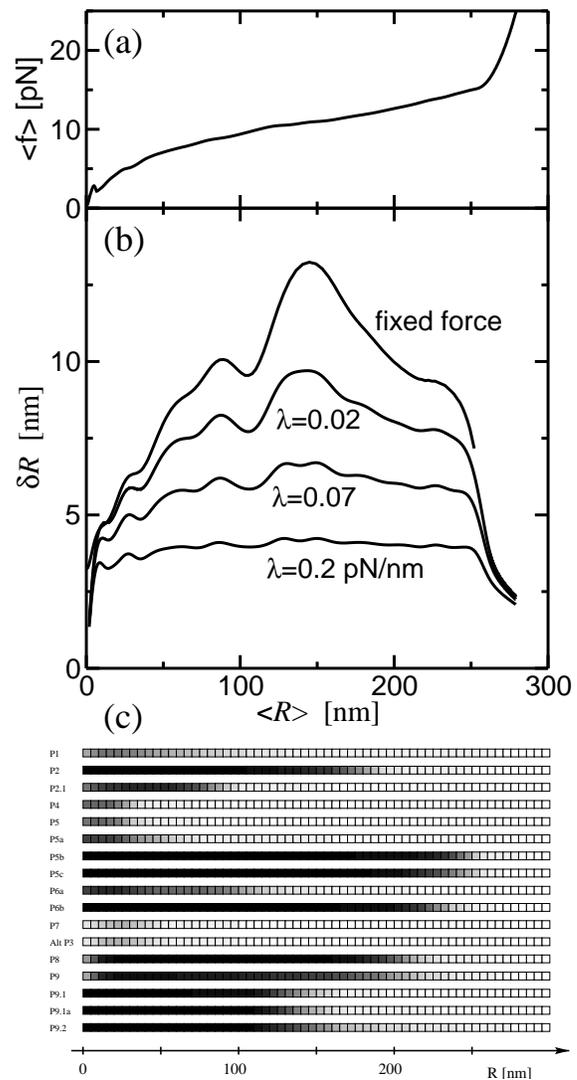}
  \caption{\label{figdR}
  Calculated force-extension characteristics of the 
  {\it Tetrahymena thermophila} group I intron. (a) FEC. 
  (b) Fluctuations in the extension of the ribozyme for different 
  stiffness \protect$\lambda$ of the force measuring device. 
  (c) Fractional opening for all of the
  structural elements at \protect$\lambda=0.02~pN/nm$. A black square
  indicates that the structural element is completely intact while a
  white square indicates that all base pairs of the element have been
  opened. It can be seen that the FEC lacks all structural
  features. In contrast, the fluctuations display pronounced peaks
  that coincide with the disappearance of specific structural
  elements.}
\end{figure} 

The quasi-equilibrium FEC of the intron is shown in Fig.~\ref{figdR}(a). 
Here, we have not added an additional linker in the calculation, since the 
intron is already fairly large and flexible, so that the effect of a 
comparatively short and stiff double-stranded linker is negligible. 
As expected, the FEC displays no signatures of the secondary structure, mostly 
due to `compensation' between different structural elements \cite{us}. 
As for the P5ab hairpin above, we compute the equilibrium fluctuations
$\delta R$, which would correspond to the average standard deviation in 
an infinitely long experimental time trace $R(t)$. 
Obviously, our averaged $\delta R$ contains less information than 
experimental $R(t)$ traces. 
Nevertheless, as Fig.~\ref{figdR}(b) shows, already the equilibrium 
fluctuations $\delta R$ display interesting sequence-dependent features, 
i.e. local maxima and minima. 
The minima can again be interpreted as intermediate states along the 
(un)folding pathway, and the maxima as structural transitions. 
Note that the central peak close to $\langle R\rangle=150$~nm splits 
into two peaks as the stiffness of the spring is increased, indicating 
the appearance of a new intermediate on the (un)folding pathway. 
This illustrates again how the ability to resolve intermediate states depends 
on the stiffness of the force-measuring device. For the choice of the 
latter, a trade-off between resolution of intermediate structures and 
fluctuation amplitude has to be respected. 

Within our theoretical model it is a simple matter to identify the 
structural transitions associated with the peaks in Fig.~\ref{figdR}(b). 
To obtain a measure for the likelihood to find a given structural element 
at a given mean extension of the molecule, we have summed the basepair 
probabilities for all basepairs in the element and divided by the total 
number of basepairs in the intact element. Here, the basepair probabilities 
at a given extension are computed using Eqs.~(\ref{xm}--\ref{Pij}). 
For each of the labeled structural elements in Fig.~\ref{figstruct}, we 
plotted the normalized total basepair probabilities as a function of 
the mean extension at $\lambda=0.02$~pN/nm in Fig.~\ref{figdR}(c) using 
a greyscale code (black corresponds to presence of the element with 
probability one and white to absence of the element). 
We observe that the opening of most structural elements is localized to 
a relatively small extension interval, however typically the opening of 
several elements takes place simultaneously. 
Roughly, the structural elements open in the following order: 
P7, Alt P3, P4, P5, P5a, P1, P2.1, P6a, P9.1, P9.1a, P9.2, P2, P9, P8, 
P6b, P5c, and P5b. 

By simultaneous inspection of Fig.~\ref{figdR}(b) and (c), we can assign 
particular peaks in the fluctuation curve to the opening of particular 
elements. 
For instance, the central peak is associated with the parallel opening 
of P9.1, P9.1a, and P9.2, whereas the most stable elements P5b, P5c open
at the final ``hump''.  
Generally, elements that free a lot of single strand upon opening, such 
as a stem-loop structure with a large loop (e.g. P2.1), are associated with 
a distinct peak in the fluctuation curve. 
This peak reflects the large difference in extension between the open and 
closed state of the element. 

To obtain the information displayed in Fig.~\ref{figdR}(c) in the 
experiment will be more laborious, but should be possible given the known 
secondary structure. Since at every mean extension (or force) only a small 
number (up to 3) structural elements open in parallel, one can expect a small 
number of characteristic plateaus in the experimental $R(t)$ traces. The
identification of the plateaus with the corresponding elements can be
obtained by cutting the molecule at several places between the known
secondary structure elements and performing the same measurement on
subsequences. A similar approach was taken for recent
nonequilibrium experiments in the Berkeley group (S. Dumont \&
I. Tinoco Jr., private communication).  Obtaining the mechanical 
(un)folding pathway for a moderately sized RNA with known secondary 
structure therefore appears feasible by measuring the fluctuations.

\begin{figure}
  \includegraphics[width=7.5cm]{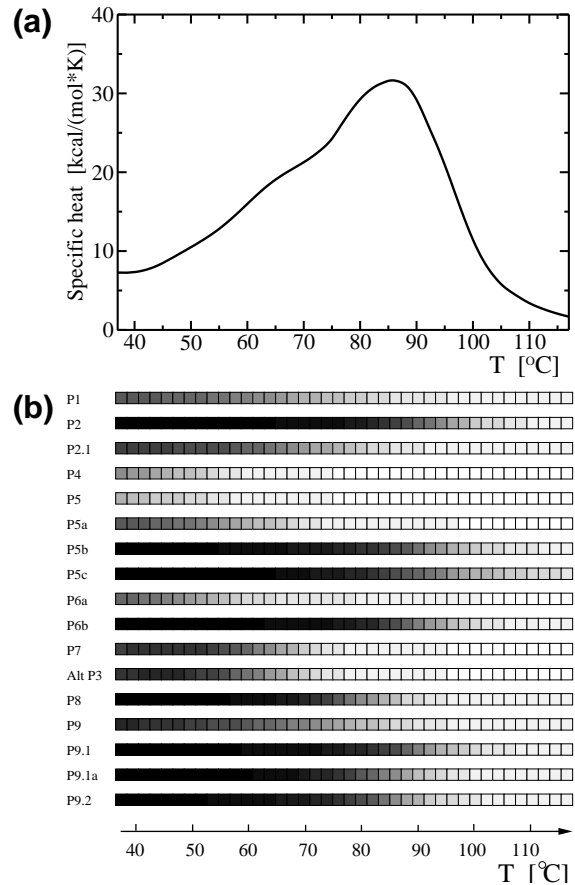}
  \caption{\label{melting}
  Thermal unfolding of the intron. 
  (a) Specific heat curve. 
  (b) Fractional opening for all of the structural elements. 
  Analogously to Fig.~\protect\ref{figdR}(c)
  a black square represents a fully formed structure whereas a white square
  represents a structure that has fully opened up.}
\end{figure} 

It is interesting to compare the mechanical unfolding studied so far
with thermal unfolding. 
The thermal unfolding process of the group I intron can be studied 
theoretically with the Vienna RNA package \cite{vienna}. 
Fig.~\ref{melting}(b) shows the fractional opening of all structural elements 
as in Fig.~\ref{figdR}(c), but plotted against temperature. 
We observe that the individual stems show a less sharp transition from 
closed to open than in Fig.~\ref{figdR}(c). 
As a result, the specific heat curve shown in Fig.~\ref{melting}(a) 
displays only a single broad peak for the unfolding of the secondary structure 
(the specific heat is obtained from the total folding free energy, 
$\Delta G$, through $-T\,\partial^2(\Delta G)/\partial T^2$). 
This is consistent with UV absorption experiments \cite{banerjee} that found 
only two broad peaks for the same RNA, where the low-temperature peak could 
be associated with melting of the tertiary structure and the high-temperature 
peak with the secondary structure. 

{From} Fig.~\ref{melting}(b), a rough order for thermal opening of the 
structural elements can be discerned: P4, P5, P5a, P6a, P7, Alt P3, P1, 
P2.1, P9, P8, P9.1a, P9.2, P9.1, P6b, P5b, P2, P5c. 
No structural element in this thermal (un)folding pathway is more than four 
list positions away from its place in the mechanical (un)folding pathway 
discussed above. 
This suggests that the thermal and mechanical unfolding pathways, although 
different in detail, can still be used as reasonable priors for each other. 
\section*{Summary and Discussion}
Single-molecule experiments are particularly useful when the molecule
under study can take on several conformations, e.g. folding
intermediates, because in these cases bulk measurements usually yield 
only average properties, while single-molecule measurements can
characterize each of the different conformations. 
Here, we predicted experimental signatures of such folding intermediates 
in the quasi-equilibrium fluctuations of the end-to-end distance. 
We found that a local minimum in the fluctuations curve indicates a 
locally stable intermediate state in the average mechanical unfolding
pathway (which is well-defined and identical to the (re)folding pathway). 
The number of local minima in the fluctuations curve can therefore serve 
as a lower estimate of the number of locally stable intermediate states 
in this pathway. 
Furthermore, these signatures could be used to reconstruct the (un)folding 
pathway of an RNA molecule with a known secondary structure by cutting the 
sequence at appropriate positions and individually probing different 
substructures of the molecule. 

We also showed that the stiffness of the force-measuring device plays a 
crucial role in determining the resolution of the quasi-equilibrium 
fluctuations curve. 
If the device is too soft, some local minima can be lost, while a too rigid 
device has a low fluctuation amplitude and therefore a low 
signal-to-noise ratio. 
The optimal choice for the stiffness (see `Results') leads to the best 
estimate for the number of locally stable intermediate states along the 
pathway and increases the degree to which the pathway could be 
reconstructed with the ``cutting approach''. 
The dependence on the stiffness has a simple intuitive explanation within 
the statistical mechanics problem of a particle with an attached spring in 
a one-dimensional energy landscape (see Fig.~\ref{stickslip}): 
A stiff force-measuring device corresponds to a rigid spring, which tends to 
localize the particle in valleys of the landscape (i.e., locally stable 
intermediates) and forces it to make small jumps as it is pulled along the 
landscape. 
On the other hand, a soft spring allows the particle to spread out and make 
long jumps, effectively hiding the intermediates in the equilibrium curve.  

We hope that this view, which also explains the smoothness of 
quasi-equilibrium FEC's for large RNA's \cite{us}, will be useful for the 
design and interpretation of future experiments.
Of course, experiments typically record not only the average fluctuations, 
but the detailed time-traces of the end-to-end distance \cite{liphardt}. 
These traces probe the kinetics of structural rearrangements in RNA, and could 
provide even more detailed resolution of intermediate states (if the kinetics 
is slow enough to be resolvable in the experiment). 
However, we expect that it is still helpful to make the choice for the 
stiffness of the force-measuring device as discussed above, in order to limit 
the number of states that contribute to each of the timetraces. 

Although single-molecule experiments on RNA have already produced remarkable 
results, there are many desirable future developments. 
For instance, it appears that the current approaches are not well-suited to 
measure the secondary structure of an unknown molecule (e.g., in the 
'cutting approach' one would not know where to cut and it would be too 
laborious to try all positions). 
Is there a direct way to mechanically measure the secondary structure? 
Also, could one detect signatures of pseudoknots or even tertiary 
interactions?
We hope that theoretical models of the type presented here can be useful 
for the planning and design of new experimental approaches in the future.

{\bf Note.} 
After completion of this work, we learned of a related theoretical 
study by S.~Cocco, J.F.~Marko, and R.~Monasson (preprint, cond-mat/0207609). 
These authors present a kinetic model for the end-to-end distance 
fluctuations of stretched RNA molecules, which nicely complements our 
thermodynamic study. 
Very recently, A.F.~Sauer-Budge, J.A.~Nyamwanda, D.K.~Lubensky, and D.~Branton 
(preprint cond-mat/0209414) have experimentally studied the unzipping 
kinetics of double-stranded DNA with a short internal loop, and demonstrated 
the presence of an intermediate state comparable to the one discussed here 
for the modified P5ab hairpin.

{\bf Acknowledgments.} 
We are grateful to S.M. Block, S. Dumont, H. Isambert, J. Liphardt,
D.K. Lubensky, I. Tinoco, and S.A. Woodson for stimulating discussions.
U.G. was supported in part by a fellowship from the `Deutscher Akademischer 
Austauschdienst'. 
R.B. and T.H. acknowledge support by the NSF through Grant No. DMR-9971456,
DBI-9970199, and the Beckmann foundation.
\end{document}